\documentclass[fleqn,usenatbib]{mnras}

\usepackage{newtxtext,newtxmath}
\usepackage[T1]{fontenc}

\DeclareRobustCommand{\VAN}[3]{#2}
\let\VANthebibliography\thebibliography
\def\thebibliography{\DeclareRobustCommand{\VAN}[3]{##3}\VANthebibliography}

\usepackage{graphicx}	
\usepackage{amsmath}	
\usepackage{xspace}
\usepackage{tabularx}

\newcommand\asloth{\textsc{a-sloth}\xspace}

\newcommand{\Msun}{\,\ensuremath{\mathrm{M}_\odot}}
\newcommand{\te}{\ensuremath{\tau_e}}

\title[A-SLOTH reveals the nature of the first stars]{A-SLOTH reveals the nature of the first stars}

\author[T. Hartwig et al.]{
Tilman Hartwig\textsuperscript{\href{https://orcid.org/0000-0001-6742-8843}{\includegraphics[width=2.5mm]{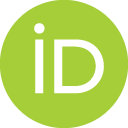}}\,}$^{1,2}$\thanks{E-mail: Tilman.Hartwig@uba.de},
Veronika Lipatova\textsuperscript{\href{https://orcid.org/0000-0002-6111-2570}{\includegraphics[width=2.5mm]{orcid.png}}\,}$^{3,4}$,
Simon C. O. Glover\textsuperscript{\href{https://orcid.org/0000-0001-6708-1317}{\includegraphics[width=2.5mm]{orcid.png}}\,}$^{3,4}$,
and Ralf S. Klessen\textsuperscript{\href{https://orcid.org/0000-0002-0560-3172}{\includegraphics[width=2.5mm]{orcid.png}}\,}$^{3,4}$ 
\\
$^{1}$Application Lab for AI and Big Data, German Environment Agency, Alte Messe 6, 04103 Leipzig, Germany\\
$^{2}$Department of Physics, School of Science, The University of Tokyo, Bunkyo, Tokyo 113-0033, Japan\\
$^{3}$Universit\"at Heidelberg, Zentrum f\"ur Astronomie, Institut f\"ur Theoretische Astrophysik, D-69120 Heidelberg, Germany\\
$^{4}$Universit\"at Heidelberg, Interdiszipli\"ares Zentrum f\"ur Wissenschaftliches Rechnen, D-69120 Heidelberg, Germany\\
}

\date{Accepted 2024 October 1. Received 2024 September 29; in original form 2024 July 27}

\pubyear{2024}

\begin{document}
\label{firstpage}
\pagerange{\pageref{firstpage}--\pageref{lastpage}}
\maketitle

\begin{abstract}
The first generation of stars (Pop~III) are too dim to be observed directly and probably too short-lived to have survived for local observations. Hence, we rely on simulations and indirect observations to constrain the nature of the first stars.
In this study, we calibrate the semi-analytical model \asloth (Ancient Stars and Local Observables by Tracing Halos), designed for simulating star formation in the early Universe, using a likelihood function based on nine independent observables. These observables span Milky Way-specific and cosmologically representative variables, ensuring a comprehensive calibration process. This calibration methodology ensures that \asloth provides a robust representation of the early Universe's star formation processes, aligning simulated values with observed benchmarks across a diverse set of parameters. The outcome of this calibration process is best-fit values and their uncertainties for 11 important parameters that describe star formation in the early Universe, such as the shape of the initial mass function (IMF) of Pop~III stars or escape fractions of ionizing photons. Our best-fitting model has a Pop~III IMF with a steeper slope, d$N$/d$M \propto M^{-1.77}$, than the log-flat models often proposed in the literature, and also relatively high minimum and maximum masses, $M_{\rm min} = 13.6~\Msun$ and $M_{\rm max} = 197~\Msun$. However, we emphasize that the IMF-generating parameters are poorly constrained and, e.g., the IMF slope could vary from log-flat to Salpeter.
We also provide data products, such as delay time distribution, bubble size distributions for ionizing and metal-enriched bubbles at high redshift, and correlation plots between all 11 input parameters. Our study contributes to understanding the formation of early stars through \asloth, providing valuable insights into the nature of Pop~III stars and the intricate processes involved in the early Universe's star formation.
\end{abstract}




\section{Introduction}
The formation of the first stars is an event of huge significance in the early history of our Universe, marking the end of the Cosmic Dark Ages and the birth of the luminous, highly structured cosmos that we observe at the present day \citep{glover05,greif15,klessen19,hammerle20,klessen23}. The first stars, commonly referred to as Population III or Pop III stars, influence their surroundings in a variety of ways, ranging from the production of ionizing and photodissociating photons to the enrichment of the gas with heavy elements ("metals"). Understanding the nature of the first stars and their impact on their environment is therefore crucial for understanding the context in which the earliest galaxies formed. Compact remnants from Pop III stars may also play an important role in the later history of the Universe, acting as the seeds from which supermassive black holes eventually grow \citep{woods19}, and contributing to the population of gravitational wave sources observed in the local Universe \citep{kinugawa14,hartwig16,liu20a}.
 
 Unfortunately, the study of Pop III stars is greatly hampered by the intrinsic difficulties involved in observing them. Pop III stars form in significant numbers only at redshifts $z > 5$, with the majority forming at $z \sim 15$--20 \citep{bromm02,yoshida03,magg16,jaacks19,skinner20,schauer20,kulkarni20,liu20b,hartwig22,hegde23}. From our viewpoint, they form at immense distances from us, so even the brightest individual Pop III stars are unobservable with current facilities unless their brightness is strongly boosted by gravitational lensing. Massive clusters of Pop III stars may be detectable, but the fraction of Pop III star formation that occurs in clusters of sufficient mass remains highly uncertain, making it hard to quantify our likelihood of finding any such clusters \citep{riaz22}.

 Progress with modelling the formation of Pop III stars with numerical simulations has also been difficult, owing the broad dynamical range of the problem and the number of different physical processes that are involved. Simulations broadly agree that Pop III star formation is highly clustered and that the Pop III stellar initial mass function (IMF) is top-heavy when compared to the present-day IMF \citep{clark11,greif15,susa19,liao19,wollenberg20,sharda20,sugimura20}. However, current simulations that follow the full duration of the Pop III star formation process in a given protogalaxy can only do so by compromising on spatial resolution, meaning that their results are not numerically converged. Conversely, simulations with sufficient spatial resolution to yield converged results can follow Pop III star formation for only a relatively short period of time. As a result, key quantities such as the Pop III star formation efficiency and the Pop III IMF remain only poorly constrained by simulations.

 One way of avoiding the problems associated with studying Pop III star formation in situ is to look for its imprints on larger scales or in the local Universe. For example, the Thomson scattering optical depth inferred from the cosmic microwave background (CMB) constrains the efficiency of Pop III star formation, while details of the low metallicity end of the Milky Way's metallicity distribution function (MDF) provide constraints on high redshift metal production by Pop III supernovae \citep{Salvadori07,Frebel15,Ishigaki18,koutsouridou23}. However, to fully exploit these observables and to correctly interpret what they can tell us about the high redshift Universe, it is necessary to compare them to the predictions of models for both the global history of Pop III star formation and the history specific to the progenitors of the Milky Way. Semi-analytic galaxy formation models (SAMs) provide a convenient framework for making these comparisons, as they allow a large volume of parameter space to be explored for moderate computational cost. In a previous paper (\citealt{hartwig22}, hereafter Paper I), we introduced A-SLOTH, a publicly available SAM specifically designed to connect the formation of the first stars and galaxies to local observables. 

In that paper, we used a set of six observables -- the CMB Thomson scattering optical depth, the stellar mass of the Milky Way, the stellar mass function of Milky Way satellite galaxies, the fraction of extremely metal poor (EMP) stars in the Milky Way's halo, the ratio of ultra metal poor (UMP) to EMP stars, and the cosmic star formation rate density at high redshift -- to calibrate the free parameters in the A-SLOTH model, and presented a set of parameter values that produced model predictions in good agreement with these observables. However, computational restrictions limited our exploration of parameter space in Paper I, and our approach there precluded any examination of the posterior distribution of the parameters. In this paper, we revisit the issue of calibrating the A-SLOTH model. We consider a broader range of observables than in Paper I and explore a much larger set of combinations of parameters, using a Markov Chain Monte Carlo (MCMC) approach to examine the effects of varying multiple parameters simultaneously. We present in this paper not only a new set of best-fitting A-SLOTH model parameters but also a discussion of the posterior distribution of these parameters.

The structure of our paper is as follows. In Section~\ref{sec:method}, we introduce the \asloth model parameters that we are trying to constrain and the observable quantities that we use for this purpose. We also outline the details of the MCMC-based method that we use to constrain the parameters. In Section~\ref{sec:results}, we present the results of our calibration. We report the best fitting values for the model parameters and discuss how they compare to the results of Paper I. With the help of the posterior distribution, we also examine how well the parameters are constrained by the observations considered in this study. 

\section{Methodology}
\label{sec:method}
In this section, we first introduce the \asloth semi-analytical model (Section~\ref{sec:asloth}) and the observables we use to calibrate it (Section~\ref{sec:obs}). We then discuss the calibration procedure in Sections \ref{sec:initialSelection} and \ref{sec:MCMC}.

\subsection{A-SLOTH}
\label{sec:asloth}
The semi-analytical model \asloth simulates the formation of stars based on dark matter merger trees \citep{hartwig22,magg22b,magg22c}. On top of this dark matter skeleton, \asloth traces baryonic contents (hot/cold gas, metals, stars), as well as radiative, chemical, and mechanical feedback. The model is calibrated based on various observables, as explained below.

\asloth is computationally very efficient and therefore the only public model that allows one to simulate the formation of individual Pop~III stars in a cosmologically representative box. \asloth can be applied to dark matter merger trees that are generated based on the Extended Press-Schechter formalism, or to merger trees extracted from dark matter simulations. The latter mode of operation has the advantage that it contains spatial information about the distribution of halos, which is used to improve the feedback model. In this study, we use dark matter merger trees extracted from two different simulations. 

For Milky Way-like halos, we use 30 merger trees from the Caterpillar project \citep{Griffen18}. These merger trees are appropriate to use when comparing \asloth predictions with observational data specific to the Milky Way, such as the stellar metallicity distribution function. However, the cosmological volumes simulated in the Caterpillar project are, by design, not representative of a randomly chosen volume. Therefore, the Caterpillar merger trees are not appropriate to use when comparing with cosmological-scale observables, such as the redshift evolution of the star formation rate density. To produce results that can be meaningfully compared with these observables, we use merger trees extracted from a simulation box of sidelength 8\,Mpc/h from \citet{Ishiyama16}. All these merger trees resolve minihalos sufficiently well and are therefore suited to track Pop~III stars.

The 11 input parameters of \asloth\ are listed in Table~\ref{tab:parameters}.
\begin{table*}
	\centering
	\caption{List of free parameters in \asloth. The first 9 entries show parameters that we also explored in \citet{hartwig22} and the last two parameters are newly added as calibratable parameters in this version. The column ``old values'' refers to the previous calibration in \citet{hartwig22}.}
	\label{tab:parameters}
	\begin{tabular}{lccccccc} 
		\hline 

		Parameter & Description &\multicolumn{1}{c}{\begin{tabular}{@{}c@{}}Initial\\range\end{tabular}}& \multicolumn{1}{c}{\begin{tabular}{@{}c@{}}MCMC\\prior\end{tabular}} & \multicolumn{1}{c}{\begin{tabular}{@{}c@{}}MCMC\\jump scale\end{tabular}} & \multicolumn{1}{c}{\begin{tabular}{@{}c@{}}Old\\value\end{tabular}} & \multicolumn{1}{c}{\begin{tabular}{@{}c@{}}Best\\Fit\end{tabular}} & Central 68\% \\ 

		\hline
  
        $M_\mathrm{max}$ & Max. mass of Pop~III stars, \Msun & $100-400$& $100 - 400$ & $0.1$\,dex & $210$ & $197$ & $110-313$\\
        
        $M_\mathrm{min}$ & Min. mass of Pop~III stars, \Msun & $0.8-10$ & $0.8-25$ & $2$ & $5$ & $13.6$ & $6.6-21.1$  \\
        
        $\alpha_\mathrm{III}$ & power-law index of the Pop~III IMF & $0-2.5$ & $-2.3$ - $2.3$ & $0.23$ & $1$ & $1.77$ & $0.23-2.27$  \\
        
        $\eta _{\rm III}$ & Pop~III star formation efficiency & $0.03-5.0$ & -- & $0.4$\,dex & $0.38$ & $8.15$ & $0.60-87.6$ \\
        
        $\eta _{\rm II}$ & Pop~II star formation efficiency & $0.03-1.0$ & -- & $0.2$\,dex & $0.19$ & $0.237$ & $0.099-1.64$ \\
        
        $\alpha_\mathrm{out}$ & slope of outflow efficiency & $0.5-2.0$ & $<5$ & $0.2$ & $0.86$ & $2.59$ & $1.78-4.05$ \\
        
        $M_\mathrm{out0}$ & outflow efficiency normalization, \Msun & $10^{9.5-10.5}$ & -- & $0.1$\,dex & $7.5 \times 10^{9}$ & $8.39 \times 10^{9}$ & $(6.22-10.92) \times 10^{9}$ \\
        
        $f_\mathrm{esc,II}$ & Pop~II ion. photon escape fraction & $0-0.7$ & $<1$ & $0.1$ & $0.6$ & $0.175$  & $0.093-0.279$\\
        
        $f_\mathrm{esc,III}$ & Pop~III ion. photon escape fraction & $0-1$ & $<1$ & $0.1$ & $0.37$ & $0.525$ & $0.196-0.865$ \\
        
        \hline
        
        $v_\mathrm{sv} / \sigma_\mathrm{sv}$ & MW streaming velocity & $0-2.2$ & $0-2.2$ & 0.1 & $0.8$ & $1.75$ & $1.18-2.16$ \\
        
        $c_\mathrm{ZIGM}$ & IGM metallicity clumping factor & $1 - 10$ & -- & $0.04$\,dex & -- & $3.32$ & $2.87-3.72$ \\
        
		\hline
	\end{tabular}
\end{table*}
The main goal of this paper is to find suitable values and reliable ranges for these input parameters, which quantify the nature of the first stars and other effects related to star formation in the early Universe. Briefly, our input parameters are the following quantities: 
\begin{enumerate}
\item The {\bf minimum stellar mass} ($M_{\rm min}$), the {\bf maximum stellar mass} ($M_{\rm max}$), and the {\bf power law index} ($\alpha_{\rm III}$) of the Pop~III IMF, which we assume can be written as
\begin{equation}
\frac{dN}{dM_{\rm star}} \propto M_{\rm star}^{-\alpha_{\rm III}}
\end{equation}
for stellar masses between $M_{\rm min}$ and $M_{\rm max}$.

\item The {\bf Pop~III star formation efficiency} ($\eta_{\rm III}$) and {\bf Pop~II star formation efficiency} ($\eta_{\rm II}$). Both of these star formation efficiencies are defined as the fraction of cold gas that converts into stars per freefall time, where the freefall time is calculated for the mean cold gas density. With this definition, it is possible to have values for the star formation efficiencies that are $>1$. Physically, these correspond to cases in which the characteristic timescale for star formation in the cold gas is shorter than the mean freefall time.

\item The {\bf outflow efficiency slope} $(\alpha_\mathrm{out})$ and {\bf normalization} ($M_{\rm out0}$). These parameters control the efficiency with which gas can be ejected from a halo due to stellar feedback. As discussed in more detail in \citet{hartwig22}, in \asloth we assume that the injection of supernova feedback energy $\Delta E_{\rm SN}$ into the hot gas results in the loss of a fraction
\begin{equation}
f_{\rm out} = \frac{\Delta E_{\rm SN}}{E_{\rm bind, hot}} \frac{1}{\gamma_{\rm out}}
\end{equation}
of the hot gas from the halo (limited to a maximum value of 1, corresponding to the loss of all of the available hot gas), where $E_{\rm bind, hot}$ is the gravitational binding energy of the hot gas and $\gamma_{\rm out}$ is an efficiency factor given by
\begin{equation}
\gamma_{\rm out} = \left(\frac{M_{\rm h}}{M_{\rm out0}} \right)^{\alpha_{\rm out}},
\end{equation}
where $M_{\rm h}$ is the mass of the halo.

\item The {\bf ionizing photon escape fraction} for Pop III stars ($f_{\rm esc, III}$) and Pop II stars ($f_{\rm esc, II}$). These values quantify the fraction of ionizing radiation emitted by these populations of stars that manage to escape into the IGM. For simplicity, we adopt a single representative value for each population, but we allow these values to vary independently. 
\end{enumerate}

In this paper, we also include two additional input parameters that were not considered in our previous study:
\begin{enumerate}
\item The {\bf baryonic streaming velocity} (i.e.\ the initial bulk velocity of the baryons with respect to the dark matter; see \citealt{Tseliakhovich10}), expressed here in units of the redshift-dependent rms value ($v_\mathrm{sv} / \sigma_\mathrm{sv}$). This was included in \citet{hartwig22} with a fixed value of 0.8, the most likely value to be found in a randomly selected volume \citep{schauer21}. However, this value is not necessarily representative of the environment in which the Milky Way formed \citep{uysal23}, and so in our present study we include the value of $v_\mathrm{sv} / \sigma_\mathrm{sv}$ for the Milky Way formation environment as a free parameter.

\item The {\bf clumping factor for the IGM metallicity} ($c_\mathrm{ZIGM}$), a new free parameter that was not included in the previous version of \asloth.
Once metal-enriched gas has been ejected from a halo, this gas can be re-accreted onto the halo via smooth accretion. In the previous version of \asloth, we assumed that the metals are homogeneously distributed in the outflowing bubble. However, in reality, homogeneous mixing is unlikely given the large size of the bubbles \citep{Ritter15}. Making this assumption therefore leads to us underestimating the metallicity of the re-accreted gas, which in turn makes it difficult to reproduce metallicity-based observables \citep{chen22b}. To mitigate this problem, we assume that the metallicity of the re-accreted gas is higher than the mean metallicity of the bubble by a factor $c_\mathrm{ZIGM} > 1$. The value of $c_{\rm ZIGM}$ is a free parameter that will be set by comparison with observations, as described in the next section. We justify this procedure in two ways. First, simulations suggest that the metallicity of hot, outflowing bubbles around early protogalaxies tends to decrease as one moves away from the protogalaxy \citep[see e.g.][]{Ritter15,magg22a}, implying that the circumgalactic medium (from which the re-accreted gas mainly comes) has a higher metallicity than the average IGM. Second, the re-accretion can be clumpy and such clumps (denser, colder pockets of gas) usually also have higher metallicities.
\end{enumerate}


As well as including these two new free parameters, we have also made two other major improvements to \asloth compared to the initial public release described in \citet{hartwig22} and \citet{magg22b}. 
\begin{enumerate}
\item We fixed a bug in the treatment of external ionizing feedback. \asloth assumes that external ionizing radiation will prevent star formation in halos that have a virial temperature below $\sim 10^4$\,K. However, in the first release, we did this check only once, at the point at which the halo is first exposed to an external radiation field, and did not check whether this remained the case at later times. In the case where the halo is rapidly growing in mass, this could result in suppressed star formation in halos with $T_{\rm vir} > 10^{4}$~K that should be immune to external ionizing feedback.
This problem has been fixed in the latest version of the code. In practice, we find that in most cases, the number of affected halos is negligible, and so the effect of this bugfix is therefore only minor (see Sec. \ref{sec:caveats}). All results in this paper are shown for the new (bug-free) version, except results of the calibration process (MCMC) itself (Sec. \ref{sec:calib}).


\item Previously, we used simple mass-based fitting functions for stellar parameters, such as stellar lifetime or the production rate of ionizing photons. Moreover, we only discriminated between Pop~III and Pop~II stars, without taking into account the specific metallicity or stellar ages. In the latest version of the code, we have replaced these fitting functions with table-based lookups using tables of stellar properties as a function of zero-age main sequence mass, metallicity, and stellar age. These tables were generated using the SEVN model \citep{SEVN19,SEVN23}. See \citet{klessen23} for a more detailed comparison of different stellar models of Pop~III stars.
\end{enumerate}

\subsection{Observables}
\label{sec:obs}
To calibrate \asloth, we maximize a likelihood function that is based on nine independent observables: the optical depth to Thomson scattering, the cosmic star formation rate density (SFRD) at low redshift, the cosmic SFRD at high redshift, the MW stellar mass, the MW metallicity distribution function (MDF), the fraction of EMP stars in the MW halo, the non-detection of any Pair-Instability Supernova (PISN) signature in EMP stars, the stellar masses of MW satellite galaxies, and the metallicities of MW satellite galaxies. In the next sections, we will explain the individual observables, how we predict them with \asloth, and how we calculate the individual likelihoods $\mathcal{L}_i$, which we eventually multiply to obtain the total likelihood.

For an observed value $x_\mathrm{obs,i}$, an uncertainty $\sigma_i$, and a simulated value $x_\mathrm{sim,i}$, the likelihood is given by
\begin{equation}
    \mathcal{L}_i(x_{\mathrm{obs,}i},\sigma_i,x_{\mathrm{sim,}i}) = \frac{1}{\sqrt{2 \pi \sigma _i ^2}} \exp \left( -\frac{(x_{\mathrm{obs,}i} - x_{\mathrm{sim,}i})^2}{2 \sigma _i ^2} \right),
\end{equation}
under the assumption that the uncertainty is normally distributed.

The observables fall into two categories: MW-specific and cosmologically representative. The cosmologically representative variables are calibrated based on cosmologically representative DM merger trees, which are extracted from a computational box of side length 8~Mpc/h, which reaches down to $z=4.5$ \citep{Ishiyama16}. The MW-specific observables are calculated for 30 independent MW-like merger trees, which are taken from the Caterpillar simulation suite \citep{Griffen18}. For each of the MW-specific observables, we calculate a mean value, $x_{30}$, and standard deviation, $\sigma_{30}$, over the set of 30 merger trees. We then require that the mean value should be close to the observed value and we add $\sigma_{30}$ quadratically to the observational uncertainty.

The final likelihood function is given by
\begin{equation}
\begin{aligned}
    \mathcal{L} = &\mathcal{L}_\tau \times \mathcal{L}_\mathrm{SFRD,HST} \times \mathcal{L}_\mathrm{SFRD,JWST} \times \mathcal{L}_\mathrm{Mstar} \times \mathcal{L}_\mathrm{MDF}^3 \times\\ 
    &\mathcal{L}_\mathrm{EMP} \times \mathcal{L}_\mathrm{PISN} \times \mathcal{L}_\mathrm{SatMass}^2 \times \mathcal{L}_\mathrm{SatMet}
\end{aligned}
\end{equation}
We explain the individual terms and motivate the weightings in the next sections.

\subsubsection{Ionisation history}
The optical depth of Thomson scattering quantifies the ionization history of the Universe. To calculate \te, we follow the implementation in \citet{hartwig22},
\begin{equation}
 \te(z) = c \sigma_{\rm T} n_{\rm H} \int _0 ^z \mathrm{d}z' f_e Q_{\rm ion} (z') (1+z')^3 \left| \frac{\mathrm{d}t}{\mathrm{d}z'} \right|,
\end{equation}
where $z$ is the redshift, $\sigma_{\rm T} = 0.665 \times 10^{-24} \, \mathrm{cm}^2$ is the Thomson scattering cross-section, $n_\mathrm{H}$ the cosmological mean density of hydrogen nuclei at $z=0$, $Q_{\rm ion}$ the volume filling fraction of ionized regions, and $f_e$ the number of free electrons per hydrogen nucleus in the ionized IGM. The presence of helium makes this number slightly larger than one and we assume for simplicity that
\begin{equation}
 f_e = \begin{cases}
 1+Y_p/2X_p & \mathrm{at}\ z \leq 4 \\
 1+Y_p/4X_p & \mathrm{at}\ z > 4,
 \end{cases}
\end{equation}
where $Y_p$ and $X_p$ are the primordial abundances of He and H, respectively \citep{Robertson13}. 

The value of the Thomson scattering optical depth is \citep{planck18}
\begin{equation}
    \tau = 0.054 \pm 0.007
\end{equation}
so that the corresponding likelihood is given by 
\begin{equation}
    \mathcal{L}_\tau(0.054,0.007,\te)
\end{equation}

\subsubsection{Star formation rate density}
The cosmic star formation rate density describes how many stars form per unit of time and per unit volume in the Universe. In the era of JWST, we can determine this quantity observationally to $z>10$. To calibrate \asloth, we use a compilation of independently observed values for the SFRD in the redshift range $4.5 \leq z \leq 13.3$. The lower redshift limit is given by our cosmological box, which only extends down to $z \approx 4.5$, and the upper limit is given by the JWST frontier. Specifically, we do not include JWST-based measurements of the SFRD at $z \sim 17$ by \citet{harikane22} and \citet{bouwens2023}, as these are based on a very small number of sources and hence are highly uncertain. 

\begin{figure}
\includegraphics[width=\columnwidth]{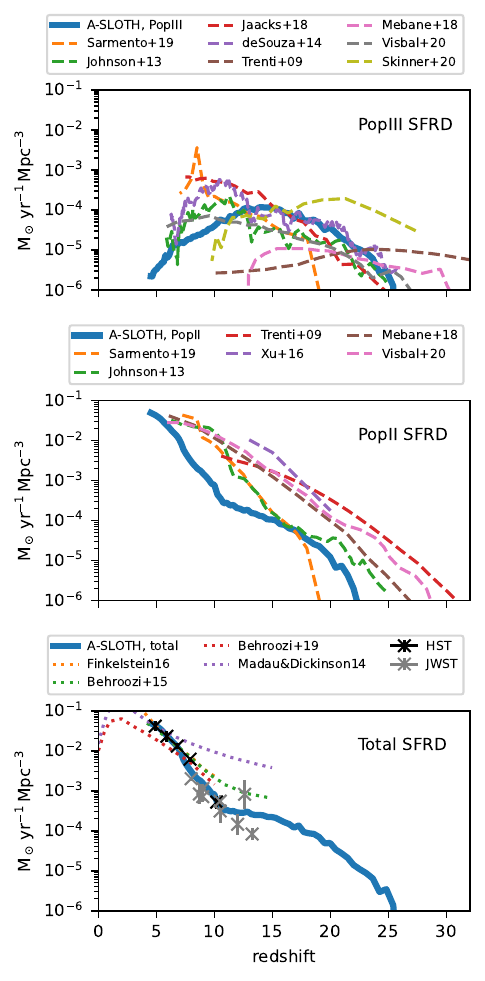}
\caption{{\it Top}: evolution of the cosmic SFRD of Pop~III stars with redshift. The blue line is the prediction from \asloth that we obtain if we use our best-fit parameters (see Section~\ref{sec:calib}). The other lines show predictions from a number of other models in the literature \citep{Trenti09,johnson13,deSouza14,Jaacks18b,Mebane18,sarmento19,Skinner2020,visbal20}. {\it Middle}: as the top panel, but for the SFRD of metal-enriched stars. For comparison, we also show predictions from a number of other models from the literature \citep{Trenti09,johnson13,Xu16a,Mebane18,sarmento19,visbal20}. {\it Bottom}: as above, but for the total SFRD (i.e.\ the sum of the Pop~III and Pop~II contributions). The blue line shows the \asloth prediction and the dotted lines show the values from the reviews of \citet{Madau14} (only valid at $z \leq 8$) and \citet{finkelstein16} and from the models of \citet{BehrooziSilk2015} and \citet{behroozi19}. The black and gray points show the values given in Table~\ref{tab:SFRD}, based on HST and JWST observations, that we aim to reproduce.}
\label{fig:SFRD}
\end{figure}

\begin{table*}
	\centering
	\caption{Compilation of SFRDs at high redshift. All values of the SFRD and its uncertainty are in $\log_{10}(\Msun\,\mathrm{yr}^{-1}\,\mathrm{Mpc}^{-3})$}
	\label{tab:SFRD}
	\begin{tabular}{lcccc} 
		\hline
		z & SFRD & $\sigma _\mathrm{SFRD}$ & Telescope & Reference \\
		\hline
4.9 & -1.387 & 0.124 & HST/ALMA & \citet{bowens16} \\
5.9 & -1.640 & 0.130 & HST/ALMA & \citet{bowens16} \\
6.8 & -1.883 & 0.076 & HST/ALMA & \citet{bowens16} \\
7.9 & -2.213 & 0.065 & HST/ALMA & \citet{bowens16} \\
10.2 & -3.287 & 0.162 & HST & \citet{oesch18} \\
\hline
8.7&-3.082 & 0.238 & JWST & \citet{bowens22} \\
10.5&-3.507 & 0.295 & JWST & \citet{bowens22} \\
12.6&-3.089 & 0.368 & JWST & \citet{bowens22} \\
8.0&-2.700 & 0.051 & JWST & \citet{donnan22,bowens22} \\
9.0&-2.945 & 0.108 & JWST & \citet{donnan22,bowens22} \\
10.5&-3.262 & 0.151 & JWST & \citet{donnan22,bowens22} \\
13.25&-4.084 & 0.151 & JWST & \citet{donnan22,bowens22} \\
9.0 & -3.147 & 0.137 & JWST & \citet{harikane22,bowens22} \\
12.0 & -3.839 & 0.26 & JWST & \citet{harikane22,bowens22} \\
		\hline
	\end{tabular}
\end{table*}
In total, we have 5 observations from HST and 9 observations from JWST (Table~\ref{tab:SFRD} and Fig.~\ref{fig:SFRD}). However, the JWST data points rely entirely on photometric redshift estimates and hence are considerably less reliable than the lower redshift measurements. Indeed, \citet{bouwens2023} have shown that the inferred SFRD at $z > 10$ can vary by up to an order of magnitude depending on the sample of candidate $z > 10$ galaxies adopted, with more robust samples yielding lower SFRDs. Moreover, the observed SFRDs are derived using an extrapolation of the luminosity functions to account for galaxies below the detection limit, which introduces another uncertainty. The values adopted here correspond to a relatively conservative sample selection, but to account for this systematic uncertainty, we weight these data points less than the HST-derived ones.


Given the measurements of the SFRD, we calculate two SFRD-based likelihoods that will enter the final total likelihood: one likelihood based on the 5 HST observations (each weighted by 1/5) and one likelihood based on the 9 JWST observations (each weighted by 1/9). The lower weighting given to the JWST data points ensures that our final likelihood is not dominated by these more numerous but less reliable points. The two terms from the SFRD that enter the total likelihood are therefore $\mathcal{L}_\mathrm{SFRD,HST}$ and $\mathcal{L}_\mathrm{SFRD,JWST}$. We calculate the likelihood for each of the observations in log-space by
\begin{equation}
    \mathcal{L}_i(SFRD_\mathrm{obs}(z),\sigma_\mathrm{SFRD} (z),SFRD_\mathrm{sim}(z)),
\end{equation}
where $\sigma_{\rm SFRD}$ is the reported one-sigma uncertainty in the SFRD, as listed in Table~\ref{tab:SFRD}.


\subsubsection{Stellar mass of the Milky Way}
\label{sec:MWmass}
For the observed stellar mass of the MW we take $M^\mathrm{obs}_\mathrm{star, MW} = (5.43\pm0.57) \times 10^{10}\Msun$ from \citet{McMillan2017}. To the observational uncertainty, we also add the statistical scatter from 30 merger trees, $\sigma_{30}$, so that the final likelihood is given by
\begin{equation}
    \mathcal{L}_\mathrm{Mstar}(5.43 \times 10^{10},\sqrt{(5.7 \times 10^{9})^2 + \sigma_{30}^2},M^\mathrm{sim}_\mathrm{star, MW}/\Msun).
\end{equation}

\subsubsection{Metallicity distribution function}
\begin{figure}
\includegraphics[width=\columnwidth]{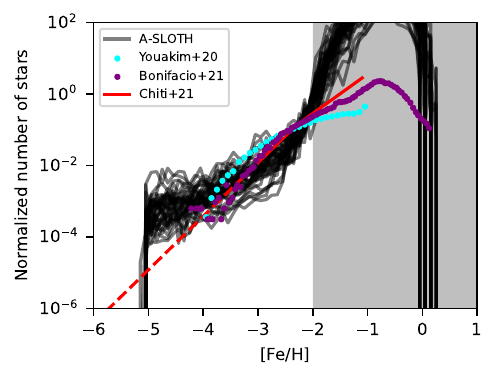}
\caption{The gray lines (which partially overlap and appear black) are the MDFs from the MW-like merger trees simulated with \asloth. The analytical shape of \citet{chiti21} is extrapolated to [Fe/H]$<-3$, where we show the red line as dashed. The observed values by \citet{Youakim20} and \citet{bonifacio21} are taken in intervals of 0.1\,dex to calculate the likelihood.}
\label{fig:MDF}
\end{figure}

The low-metallicity tail of the MDF in the Milky Way is an important tracer of the first epochs of star formation. We therefore aim at reproducing the MDF in the rage $-5 <$ [Fe/H] $< -2$. Observations are incomplete below [Fe/H]$<-5$. Above [Fe/H]$= -2$, Type Ia supernovae (SNe) start to contribute, which we do not yet model in \asloth. We use three independent observational constraints in this range \citep{Youakim20,chiti21,bonifacio21}. We calculate the likelihood, $\mathcal{L}_\mathrm{MDF}$, in bins of width $0.1$. In each bin, we first calculate the mean and scatter from the 30 MW-like realizations. Then we compare them to the three observed values so that we get three terms per MDF bin. Eventually, we sum up the log-likelihoods and divide by the total number of terms as normalization. The best fitting MDFs can be seen in Fig.~\ref{fig:MDF}.
The simulated and observed MDFs agree well in the range $-4 \lesssim $ [Fe/H] $\lesssim -2$, in which there are many observations. At lower metallicities, there is not much observational guidance and \asloth produces more UMP stars than what the extrapolation of the results from \citet{chiti21} would suggest.

Because the MDF is such an important observable for metal-poor stars and because our constraint is based on 3 independent observations, we weight the contribution of $\mathcal{L}_\mathrm{MDF}$ to the final likelihood with a weight of 3.

\subsubsection{Fraction of EMP stars}
We use the fraction of EMP stars in the MW as an additional observational constraint. To obtain an observational estimate, we take a MW stellar mass of $5.43 \times 10^{10}$~$\Msun$ (see Sec.~\ref{sec:MWmass}) and a mass of the stellar halo of $10^9\Msun$ \citep{BullockJohnston05}. Furthermore, we assume that 1 in every 800 halo stars is an EMP star \citep{Youakim20}. This yields a total EMP fraction of $\sim 10^{-4.7}$. We implicitly assume that all EMP stars simulated with \asloth end up in the stellar halo. While this might not be entirely true \citep{sestito23}, the number of EMP stars in the stellar halo is certainly dominant compared to other parts of the MW \citep{chen23}.

We note that this observed fraction of EMP stars is smaller than the fraction found by \citet{chen23} in the TNG-50 simulation of around $10^{-3}$. There are various explanations for this discrepancy, such as numerical resolution or the TNG star formation recipe at the lowest metallicities. Since there is such a large difference between the observational estimate and other simulations, we assume a rather large uncertainty on this fraction of 1\,dex.

The resulting likelihood is therefore
\begin{equation}
    \mathcal{L}_\mathrm{EMP}(-4.7,\sqrt{1.0 + \sigma_{30}^2},\log _{10} (f_\mathrm{sim,EMP/All})).
\end{equation}

\subsubsection{Non-detection of PISN signature}
Among all observed EMP stars, there is no convincing chemical signature of a PISN \citep{aguado23}. A PISN produces a characteristic odd-even pattern in its nucleosynthetic yields \citep{HegerWoosley2002,kozyreva14}. Such a pattern should also be observed in the chemical composition of EMP stars if these formed out of gas that was enriched by a PISN \citep{magg22a}. To use this non-detection as an observational constraint, we first need to formalize the criterion for a reliable non-detection.

There are claims of chemical PISN signatures in high-z gas clouds \citep{yoshii22} and in multi-enriched very metal-poor stars \citep{Salvadori19,aguado23}. Therefore, we will only consider EMP stars in the MW halo, since there has not yet been a detected PISN signature in their chemical fingerprint, as shown by \citet{koutsouridou24} who also used the non-detection of PISN to make constraints on the first stars. Moreover, we will only consider stars that are mono-enriched by a PISN. Multi-enrichment and contributions from CCSNe could dilute the odd-even pattern and hence make it more difficult to detect \citep{hartwig18a,Salvadori19,hansen20,magg20}. In \asloth, we can count the EMP stars that contain elements from exactly one PISN (see also \citealt{koutsouridou24}).

In addition, we need to know the sample size of EMP stars for which we know that they are not enriched by exactly one PISN. There are >500 observed EMP stars \citep{saga17}. To exclude that a star was enriched by a PISN, a certain combination of elements needs to be observed. According to \citet{lucey22}, the three elements Al, Ca, and Mg are sufficient to exclude enrichment from a PISN. According to the Saga database \citep{Saga}, there are 302 EMP stars for which these elements are observed. There are 33 additional stars for which at least 4 even and 2 odd elements are observed. We use this combination of $N_\mathrm{notPISN} = 335$ EMP stars as the basis for our non-detection constraint. This number is robust with respect to the exact number of odd/even elements that we require to be observed.

If there are $N_\mathrm{EMP}$ EMP stars in the MW, then there are
\begin{equation}
    N_\mathrm{all} = \binom{N_\mathrm{notPISN}}{N_\mathrm{EMP}}
\end{equation}
different ways of selecting $N_\mathrm{notPISN}$ stars, where this notation represents the binomial coefficient. However, if there are $N_\mathrm{PISN}$ EMP stars in the MW that are mono-enriched by a PISN, then there are only
\begin{equation}
    N_\mathrm{not} = \binom{N_\mathrm{notPISN}}{N_\mathrm{EMP}-N_\mathrm{PISN}}
\end{equation}
combinations of $N_\mathrm{notPISN}$ EMP stars that result in a non-detection of EMP stars that are mono-enriched by a PISN. The likelihood that we end up with no detection of a PISN signature, despite having $N_\mathrm{PISN}$ EMP stars in the MW that are mono-enriched by a PISN is therefore
\begin{equation}
   \mathcal{L}_\mathrm{PISN}  = N_\mathrm{not} / N_\mathrm{all}.
\end{equation}

\subsubsection{Satellite masses}
According to \citet{GarrisonKimmel} and \citet{Munoz2018}, there are ten Milky Way satellite galaxies above the observational completeness limit of $\sim 10^{5.5}\Msun$ in stellar mass (see also \citealt{dw20}).
To obtain a more continuous likelihood measure in contrast to a classical KS test, we use the stellar mass of the tenth most massive satellite galaxy to calibrate \asloth. The tenth most massive satellite galaxy is Ursa Minor with a stellar mass of $\sim 10^{5.6 \pm 0.1}\Msun$ \citep{Simon2018}. The uncertainty is taken as the difference between the ninth and eleventh most massive satellite galaxy. We require that the tenth most massive MW satellite that is simulated with A-SLOTH also has a mass of $\sim 10^{5.6 \pm 0.1}\Msun$.

By probing the mass of the tenth most massive satellite, we are indirectly sensitive to the mass distribution of more and less massive satellites. In addition, this quantity is continuous, in contrast to the discrete steps of the KS-test, which was previously applied.

We also add the statistical scatter from 30 merger trees, $\sigma_{30}$, to the observational uncertainty so that the final likelihood is given by
\begin{equation}
    \mathcal{L}_\mathrm{SatMass}(5.6,\sqrt{(0.1)^2 + \log(\sigma_{30})^2},\log_{10}(M^\mathrm{sim}_\mathrm{star,10th}/\Msun)).
\end{equation}
In the final sum of the log-likelihood, we weigh this term twice because we found that a stronger emphasis on the satellite stellar masses also improves the mass-metallicity relation and the satellite mass function (which we do not actively calibrate for), without negatively affecting other observables.

\begin{figure}
\includegraphics[width=\columnwidth]{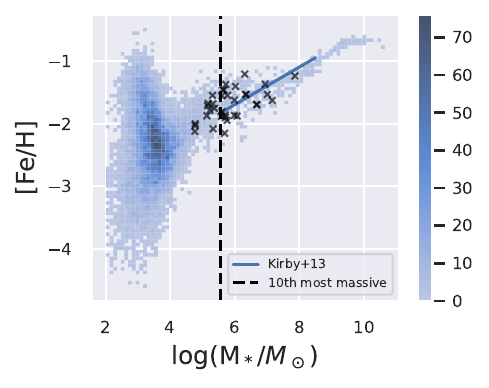}
\caption{Mass-metallicity relation for MW satellite galaxies. Each black cross is the tenth most massive satellite of each merger tree (we show 30 realizations together). The black crosses should be close to the black vertical line, which indicates the observed mass of the tenth most massive satellite. All simulated galaxies should follow the blue line \citep{kirby13} in the range over which the blue line is indicated, i.e., valid.}
\label{fig:SatMassMetal}
\end{figure}

The mass-metallicity relation together with the best-fitting model predictions are shown in Fig.~\ref{fig:SatMassMetal}.

\subsubsection{Satellite metallicity}
The mean stellar metallicity of MW satellites is very informative about star formation at high redshift \citep{salvadori15} and therefore very useful to calibrate \asloth \citep{chen22b}. To account for the observational completeness and survey selection functions, we do not directly compare satellite metallicities. Instead, we compare the mass-metallicity relation of MW satellites by \citet{kirby13}. This has the advantage that we can directly select and compare satellites above the observational completeness limit.

Based on \citet{kirby13}, we use the relation
\begin{equation}
    \mathrm{[Fe/H]} = (-1.69 \pm 0.04) + 0.30 \log _{10} \left( \frac{M_*}{10^6\Msun} \right),
\end{equation}
between mean stellar metallicity and stellar mass of MW satellites, where the uncertainty represents the rms scatter.

The observation of MW satellites is complete above a stellar mass of $\sim 10^{5.5}\Msun$ \citep{GarrisonKimmel,Munoz2018}. We compare them up to $10^{8.5}\Msun$, because this is the mass range over which the fit was fitted and it is close to the mass of the Small Magellanic Cloud ($10^{8.65}\Msun$), which might be dominated by Type Ia SNe (an efficient source of iron), which we do not include in A-SLOTH.

For each simulated MW satellite above the completeness limit, we calculate its likelihood as
\begin{equation}
    \mathcal{L}_\mathrm{SatMet}([\mathrm{Fe/H}]^\mathrm{obs}(L^\mathrm{sim}), 0.16, [\mathrm{Fe/H}]^\mathrm{sim},
\end{equation}
where $0.16$ is the rms error of the fit in \citet{kirby13}. Eventually, we add the logarithms of these likelihoods and divide by the total number of satellites that entered this calculation as a weighting factor.


\subsection{Initial model selection}
\label{sec:initialSelection}
We need to find the optimal model configuration in 11 dimensions (equal to the number of free parameters in \asloth) for a model that requires running 1+30 merger trees for each new set of parameters. Although \asloth\ is surprisingly fast for a sloth, we cannot afford a brute-force parameter exploration strategy. Therefore, we divide the optimization into two steps: the initial model selection (this section) and an MCMC exploration of the most promising models (next section).

For the initial model selection, we explore all 11 parameters simultaneously by selecting random values in the specified ranges in Table~\ref{tab:parameters}. Altogether, we ran $2^{11}=2048$ different models at this stage in order to explore sufficiently many parameter combinations in each dimension.

The goal of this initial exploration is to be able to select promising starting points for the MCMC chains in order to minimize burn-in removal. The explored parameter ranges in this initial model selection are given in column 3 of Tab.~\ref{tab:parameters}. The exact choice of these ranges is not crucial because they are only used to find an improved starting point for the MCMC exploration. However, we still motivate these choices for completeness and because some arguments are equally valid for the MCMC exploration in the next section.

The mass range for $M_\mathrm{max}$ should include the mass range in which we expect PISNe ($140-260$\Msun). The lower limit ($100\Msun$) for this upper mass is motivated by the fact that metal-free stars should be at least as massive as metal-enriched stars \citep{KroupaIMF,ChabrierIMF}. The upper limit ($400\Msun$) is chosen because Pop~III stars with $\gtrsim 300\Msun$ produce the same amount of ionizing photons per stellar baryon and do not provide any metal yields at the end of their lifetime \citep{klessen23}. Therefore, the signature of such massive Pop~III stars is indistinguishable from our likelihood-based calibration.

The lower limit for the explored range of the minimum mass ($M_\mathrm{min} \geq 0.8$) is motivated by the non-detection of any Pop~III survivors \citep{hartwig15b,Ishiyama16,magg18,magg19,Rossi2021}.

The slope of the Pop~III IMF can vary from linear-flat ($\alpha_{\rm III}\geq0$) to a bit steeper than Salpeter ($\alpha_{\rm III} \leq 2.5$). Most numerical simulations find a log-flat ($\alpha_{\rm III} = 1$) distribution of the Pop~III IMF \citep{klessen23}. Our explored range should include these results from simulations and allow some variations in both directions.

The streaming velocity is limited to $v_\mathrm{sv} / \sigma_\mathrm{sv} \leq 2.2$ because the critical halo mass for star formation via H$_2$ cooling becomes independent of the SV above this value \citep{uysal23}. Higher SVs are therefore degenerate and not distinguishable from one another. 

The other parameter ranges are based on previous experience \citep{chen22a,chen22b,hartwig22}, physical limitations (such as $f_\mathrm{esc}\leq1$), and extensive experiments. The best-fitting parameters after this random exploration are presented in Table~\ref{tab:BestFit}.

\begin{table*}
	\centering
	\caption{List of observables and their best fitting values after our initial random exploration. The best model fits correspond to the median values of the MCMC exploration and $\sigma _{30}$ are the scatter between the 30 MW-like realizations. The value of $\sigma_{30}$ for the MDF is the average value in the normalized MDF in the metallicity range $-5$ to $-2$.}
	\label{tab:BestFit}
	\begin{tabular}{lccr} 
		\hline
		Description & Observed Value & Best Fit & $\sigma _{30}$\\
		\hline
Optical Depth to Thomson scattering & 0.054 & 0.044 & -- \\
SFRD & -- & see Fig.~\ref{fig:SFRD} & -- \\
MW stellar mass & $5.43 \times 10^{10}\Msun$ & $9.3 \times 10^{9}\Msun$ & $2.8 \times 10^{10} \Msun$ \\
MDF & -- & see Fig.~\ref{fig:MDF} & 0.4 \\
EMP fraction & $10^{-4.7}$ & $10^{-5.1}$ & 0.16\,dex \\
PISN fraction & 0/335 & 0.0066 & -- \\
Satellite Masses & $10^{5.5}\Msun$ & $10^{5.8}\Msun$ & $0.7$\,dex \\
Satellite Metallicities & -- & see Fig.~\ref{fig:SatMassMetal} & -- \\
		\hline
	\end{tabular}
\end{table*}

\subsection{MCMC}
\label{sec:MCMC}
We then use the 6 best models (the ones with the highest likelihood) as starting points for MCMC chains. The number 6 is based on the computational capabilities of our cluster, which allows us to run 6 chains efficiently in parallel.

To speed up convergence, we already start at relatively good positions, found by the initial brute-force model selection. In each MCMC step, we randomly choose 3 out of the 11 free parameters, for which we vary their value, based on the MCMC jump scale (see Table~\ref{tab:parameters}). We ensure that each parameter is drawn with an equal probability so that after 11 steps, each parameter was sampled three times.

The prior ranges and distributions are given in Table~\ref{tab:parameters}. The priors on $M_{\rm max}$ and 
$v_\mathrm{sv} / \sigma_\mathrm{sv}$ are the same as in our initial model selection, for the same reasons, and the priors on $f_{\rm esc, II}$ and $f_{\rm esc,III}$ simply reflect the fact that photon escape fractions greater than one do not make physical sense. The remaining set of priors is motivated as follows:
\begin{itemize}
    \item As in our initial model selection, the minimum mass of Pop~III stars is limited at the low end by the non-detection of any Pop~III survivors \citep{hartwig15b,Ishiyama16,magg18,magg19,Rossi2021} and at the high end by the fact that some second-generation stars have been observed with the chemical fingerprint of a $\sim 25\Msun$ Pop~III SN \citep{Ishigaki18}. In our initial model selection, we conservatively required $M_{\rm min} < 10 \: {\rm M_{\odot}}$, but for the MCMC prior we simply require $M_{\rm min} < 25 \: {\rm M_{\odot}}$, to accommodate this observation.   
    \item The slope of the Pop~III IMF is limited by the Salpeter slope towards the bottom-heavy side and we limit it to its symmetric counterpart towards the top-heavy side. This is necessary because the minimum mass and maximum mass of the Pop~III IMF would become degenerate if the slope is too steep.
    \item We limit the slope of the outflow efficiency to $<5$ because the outflow model becomes insensitive to the slope for such steep values and hence the MCMC chain might diverge.
\end{itemize}
The MCMC jump scale is the width of the normal distribution (which is centered at the last parameter position) from which we draw the next parameter value to explore. If the scale is given in dex, we sample these parameters in log space. In total, we ran 10556 models in the MCMC production run.

For each of the six independent MCMC chains, we remove the first 350 data points (burn-in). After these first epochs, the standard deviation of the chains has stabilized, i.e. the standard deviations of the individual chains are similar to the standard deviation of the pooled chains \citep{gelman92}.

\begin{figure}
\includegraphics[height=0.9\textheight]{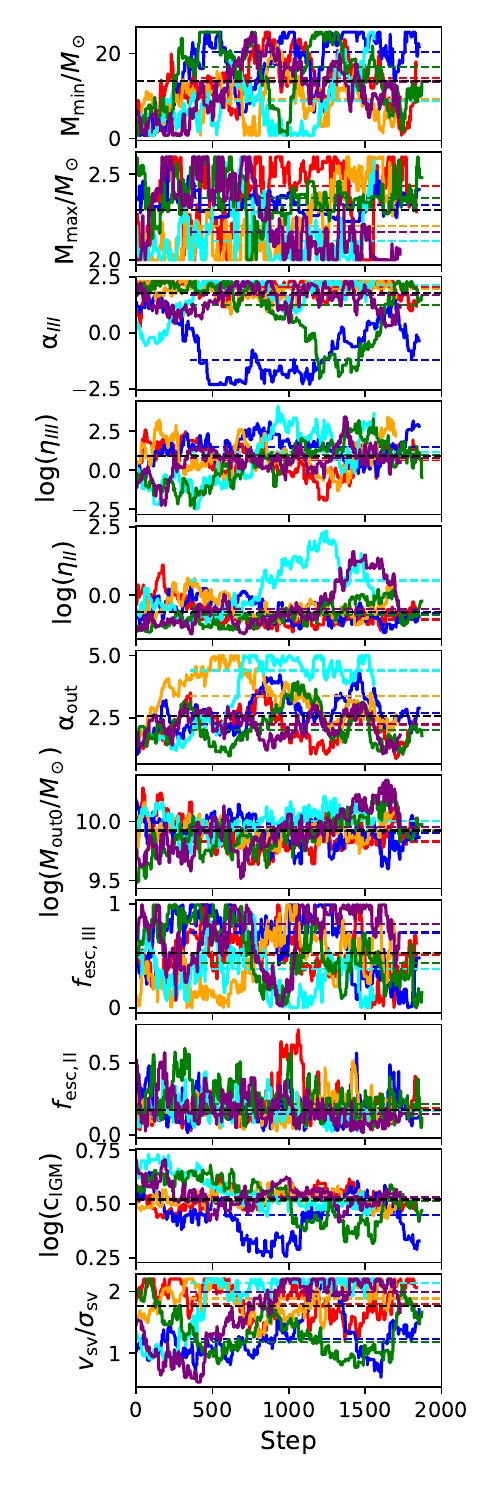}
\caption{The panels show the evolution of the MCMC chains for the 11 input parameters. Each color is one independent chain. The horizontal lines show the individual medians of the chains (after removing the burn-in) and the black line shows the median of all data points.}
\label{fig:MCMCtrace}
\end{figure}

\section{Results}
\label{sec:results}
\subsection{Calibration}
\label{sec:calib}
The trace plots for the MCMC exploration are shown in Fig.~\ref{fig:MCMCtrace}.
The 6 independent chains show a typical behavior for such a high-dimensional parameter exploration. Some chains temporarily diverge from the central mean value to explore regions of lower probability, but overall, all chains oscillate around similar mean values.

\begin{figure*}
\includegraphics[width=\textwidth]{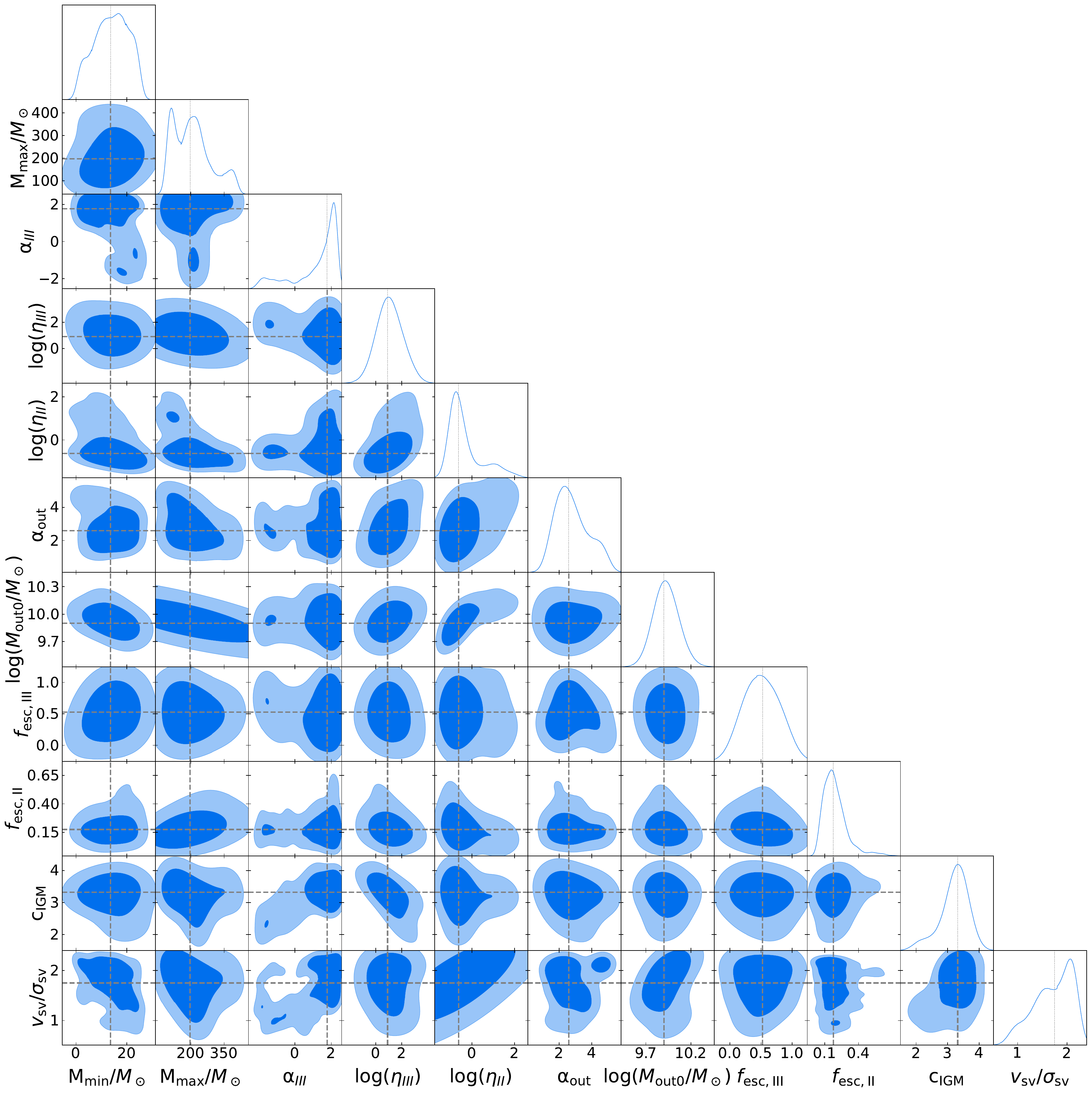}
\caption{Corner plots of our 11 explored input parameters. The dark (light) blue contour shows the 68\% (95\%) credibility range. For most parameters, the position of the median (dashed lines) and the peak of the marginalized PDFs (top of each column) agree very well. The only exceptions are the slope of the Pop~III IMF and the streaming velocity, for whose the distributions are dominated (i.e. limited) by the prior. The full data will be provided as online-only material.}
\label{fig:MCMCtriangle}
\end{figure*}

The resulting triangle plots of this MCMC exploration are shown in Fig.~\ref{fig:MCMCtriangle}.
This representation allows various conclusions. The median values of the distributions (gray dashed lines) indicate the optimal fit values of our model, which are also listed in Table~\ref{tab:BestFit}. The distribution and blue scatter contours illustrate the uncertainty of each parameter. Moreover, the contours also indicate possible correlations between input parameters. E.g., the baryonic streaming velocity and the Pop~II star formation efficiency (SFE) are positively correlated, whereas the outflowing mass normalization and the maximum mass of the Pop~III IMF are negatively correlated. This is the first inference of parameters that define the nature of the first stars that also include the full PDFs.

Looking at the best-fitting values in Table~\ref{tab:BestFit}, we find several points of immediate interest. First, we see that the best-fitting value of the Pop~III IMF slope is d$N$/d$M \propto M^{-1.77}$, somewhat steeper than indicated by recent numerical simulations \citep{klessen23}, 
but still top heavy in comparison to the present-day IMF. However, we also see from Table~\ref{tab:parameters} and Figure~\ref{fig:MCMCtriangle} that the constraint on the slope is fairly weak, with the 68\% credibility range extending down to values below 1, and so based on the observations considered here, we cannot exclude a log-flat IMF. The minimum and maximum Pop~III stellar masses are both constrained to within about 50\%. Despite the non-detection of PISN signatures being one of our observational constraints, the best-fit value of 197~M$_{\odot}$ for $M_{\rm max}$ lies within the mass range of 140--260~M$_{\odot}$ where we expect metal-free stars to end their lives as PISN, suggesting that the existence of at least some Pop III PISN cannot be ruled out on the basis of current observational data. As in Paper I, the minimum Pop~III mass favoured by the model is $\gg 1 M_{\odot}$, i.e.\ the model predicts that no Pop~III stars have masses low enough to survive on the main sequence until the present day. Note, however, that this constraint would likely weaken significantly were we to consider IMF shapes more complicated than a single power law.


The calibrated model favours a high value for the Pop~III SFE, $\eta_\mathrm{III} = 8.15$. As discussed previously, the fact that this value is greater than 1 indicates that the timescale over which Pop~III stars form in a minihalo is shorter than the mean free-fall time of the cold gas. At high redshift, in a minihalo in which the gas is cooling efficiently, the mean free-fall time is $\sim 10$~Myr, so the best-fit value of $\eta_{\rm III}$ corresponds to an absolute value for the SFE of $\sim 8 (t_{\rm SF} / 10 \: {\rm Myr})$, where $t_{\rm SF}$ is the duration of star formation in the minihalo. Numerical simulations of Pop~III star formation that account for stellar feedback typically find $t_{\rm SF} \sim 10^{4} \: {\rm yr}$ \citep[see e.g.][]{Hosokawa16}, in which case the \asloth best-fit value for $\eta_{\rm III}$ corresponds to an absolute SFE of around 1\%. For comparison, numerical simulations that follow the whole duration of the star formation process typically find SFEs of around 0.1--1\%, consistent with the \asloth value.
It should also be noted that the constraint on $\eta_{\rm III}$ provided by the model is weak: values of $\eta_{\rm III}$ as much as an order of magnitude larger or smaller than the best-fit value remain within the central 68\% of the distribution. 
The Pop~II SFE, $\eta_\mathrm{II}$, has a much smaller best-fit value, consistent with the idea that Pop~II star formation is much less bursty than Pop~III star formation, as a consequence of it taking place primarily in much larger dark matter halos.

The best-fit ionizing escape fractions ($f_\mathrm{esc,II}=0.175$, $f_\mathrm{esc,III}=0.525$) are higher than the values of a few percent found for star-forming galaxies at low redshift \citep[see e.g.][]{Khaire2016} but are in reasonable agreement with models of ionizing photon escape from high redshift minihalos \citep{Schauer2017} and galaxies \citep{lewis22,kostyuk22,trebitsch23}, which typically find high values of $f_{\rm esc}$. That said, we also see from Figure~\ref{fig:MCMCtriangle} that $f_\mathrm{esc,III}$ is only weakly constrained by current observations. The constraint on $f_\mathrm{esc,II}$ is much better. This likely reflects the fact that it is primarily photons from Population II stars that reionize the Universe, and so the Thomson scattering optical depth of the CMB is a much stronger constraint on $f_\mathrm{esc,II}$ than $f_\mathrm{esc,III}$. 
 
The optimal local value of the baryonic streaming velocity is $v_\mathrm{sv} / \sigma_\mathrm{sv} = 1.75$, which is exactly the same value that was found in \citet{uysal23} with a previous version of \asloth.

\subsection{Model predictions}

\subsubsection{Ionization History}
\begin{figure}
\includegraphics[width=\columnwidth]{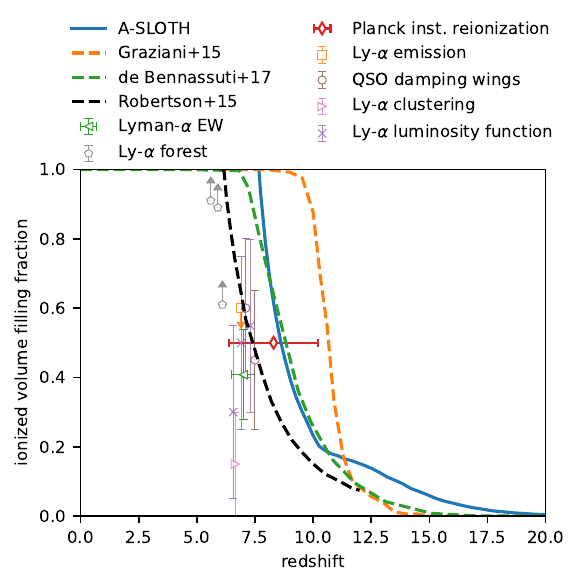}
\caption{Evolution of the ionized volume filling fraction as a function of redshift. The model prediction from \asloth (solid blue) agrees well with other models (dashed lines) and observations (points).}
\label{fig:Vion}
\end{figure}

We show the ionization history derived from our 8\,Mpc box as compared to various observations and a selection of models \citep{robertson15,graziani15,deBennassuti17} in Fig. \ref{fig:Vion}.
Our ionization history (blue line) agrees well with observational constraints \citep{McGreer15,mesinger15,Planck2016,greig17,zheng17,ouchi18,mason18,banados18,konno18} on the volume filling factor of ionized gas. \asloth predicts a higher ionized volume filling fraction than other models at about $z>12$, due to the relatively high Pop~III SFRD (see Fig.\ref{fig:SFRD}). The evolution of the ionized volume filling fraction at these redshifts is poorly constrained by current observations (although models in which it is large are ruled out by the CMB Thomson scattering optical depth), but future 21~cm observations will provide much stronger constraints here \citep[see e.g.][]{Ghara2024}.

\begin{figure}
\includegraphics[width=\columnwidth]{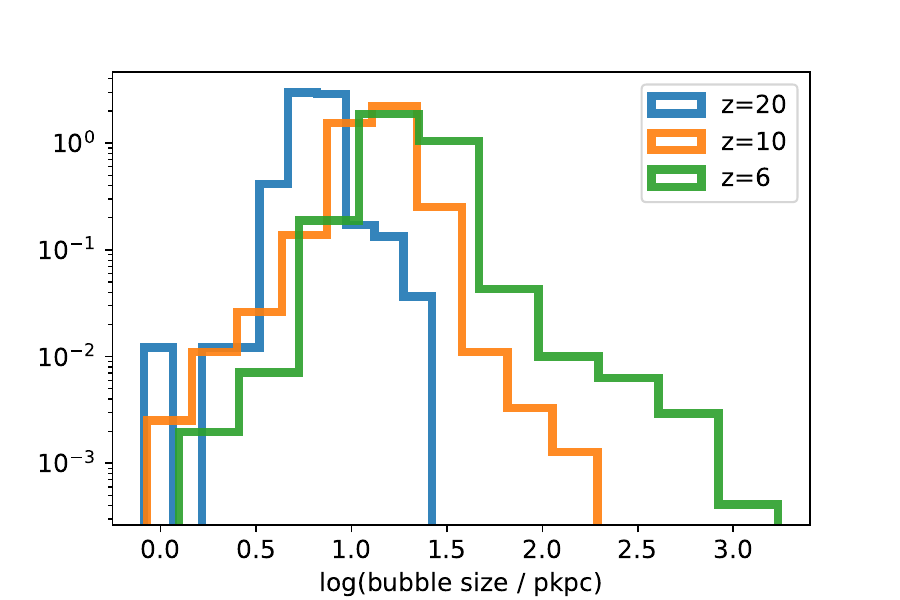}
\caption{Normalized bubble size distribution of ionized bubbles at different redshift. The sizes are shown in physical units (denoted `pkpc' in the figure).}
\label{fig:Rion}
\end{figure}

To better understand the process of reionization, we also show the bubble size distribution in Fig.~\ref{fig:Rion}.
The histograms show the distribution of the radius (in physical kpc) of the ionized bubbles at different redshifts. We can see that the bubbles grow over time, from a maximum size of about 20\,kpc at $z=20$ to a maximum bubble size of around 1\,pMpc at $z=6$. Note, however, that the values shown here assume that the bubbles can be treated as isolated objects, i.e.\ they do not account for the effects of bubble merging, which becomes increasingly important as one approaches the epoch of reionization.

\subsubsection{Luminous Satellites}
\begin{figure}
\includegraphics[width=\columnwidth]{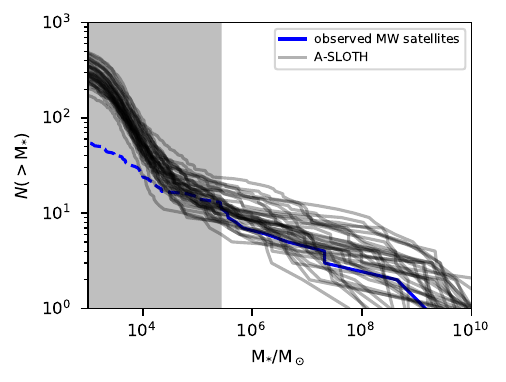}
\caption{MW satellite mass function at $z=0$. The gray lines show the satellite stellar masses from the MW-like realizations, the blue line shows the known MW satellites \citep{McConnachie2012,Munoz2018}, and the gray area (and the dashed part of the blue line) indicates the region of observational incompleteness.}
\label{fig:CumSat}
\end{figure}

Fig.~\ref{fig:CumSat} shows the MW satellite mass function.
In the stellar mass range above $M_* \sim 10^5 \Msun$, where observations are complete, \asloth reproduces the observations very well. However, \asloth predicts hundreds of luminous MW satellites below the current completeness limit.

\subsubsection{Recovery Time}
The recovery time quantifies the balance between Pop~III feedback and the ability of a halo to keep gas and/or to accrete new gas. For computational efficiency, we define the recovery time as the time between the first Pop~III SN and the first Pop~II SN that occurs later on the same merger tree branch. A long delay time means that the radiation feedback from the Pop~III stars and the mechanical feedback from the Pop~III SN was so efficient that it delayed the next generation of star formation due to the lack of fresh cold gas.

\begin{figure}
\includegraphics[width=\columnwidth]{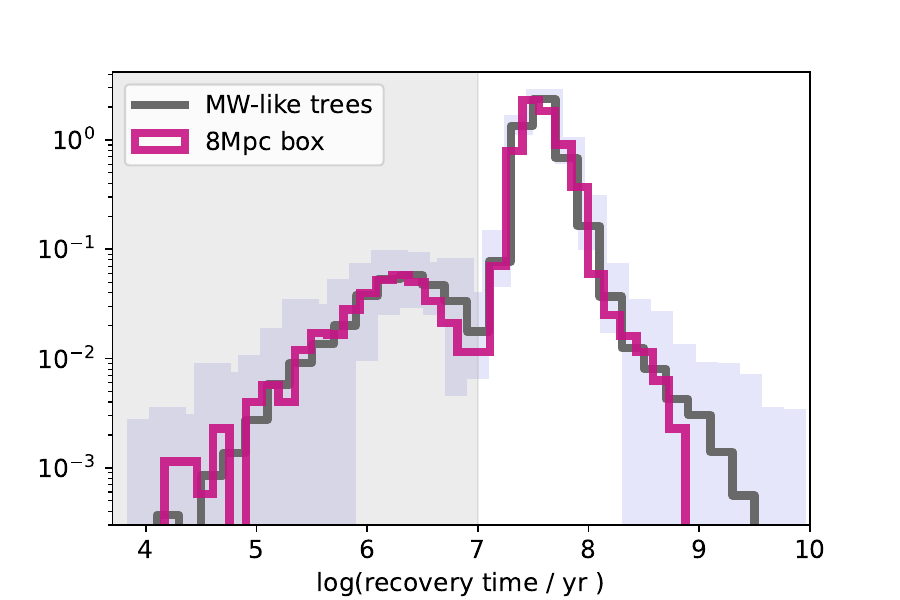}
\caption{Normalized recovery time distribution. The shaded band shows the statistical scatter from the MW realizations. The shaded area below 10\,Myr is not reliable due to limitations of the SN feedback implementation in \asloth.}
\label{fig:RecovTime}
\end{figure}

We show the recovery time between Pop~III and Pop~II star formation in Fig.~\ref{fig:RecovTime}.
The distribution of delay times shows a bimodality with one dominant peak around 30\,Myr and another smaller peak at around 3\,Myr. However, a thorough investigation of the cause of this bimodality has shown that the existence of the smaller peak is a numerical artifact. Once a Pop III star reaches the end of its life in \asloth, it ceases to act as a source of ionizing radiation. However, for reasons of computational efficiency, it only explodes as a supernova when the is some gas available in the halo. If the halo has already been cleared of gas by the prior ionizing feedback, this can cause the explosion to be artificially delayed, and can hence lead to an artificially short time delay between the Pop III SN and the first Pop II SN occurring in a given halo. The portion of the recovery time distribution that is potentially affected by this is shaded in gray in Figure~\ref{fig:RecovTime}. 

Focussing for now on the portion of the recovery time distribution that we trust, we see that we recover similar distributions for the cosmologically-representative 8~Mpc box and for the MW-like merger trees. The mean value and standard deviation of this distribution are $10^{7.5 \pm 0.3} \, \mathrm{yr}$. This range of roughly 10-100\,Myr agrees very well with values previously measured in detailed hydrodynamical simulations \citep{Jeon14, Chiaki18, latif20, Hicks2021, Magg21a}. The fact that \asloth can reproduce these values without being explicitly incentivized to do so helps increase our confidence in the robustness of the model.

\subsubsection{Chemical Enrichment History}
\begin{figure}
\includegraphics[width=\columnwidth]{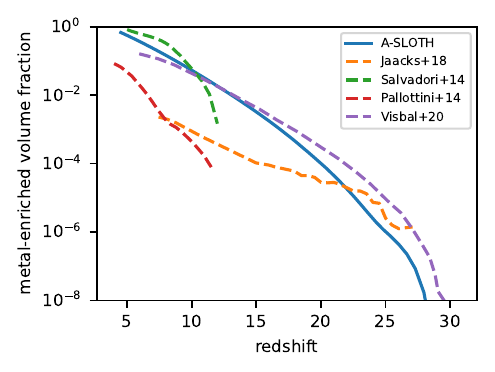}
\caption{Evolution of the metal-enriched volume filling fraction. The prediction from \asloth (blue) agrees well with other models \citep{salvadori14,pallottini14,Jaacks18b,visbal20} (dashed lines).}
\label{fig:Vmetal}
\end{figure}

The detailed treatment of Pop~III stars and their feedback in \asloth allows us to also follow the chemical enrichment history of the Universe. We show the evolution of the metal-enriched volume filling fraction in Fig.~\ref{fig:Vmetal}.
The solid blue line shows the fraction of metal-enriched volume as a function of redshift. Here, any metallicity above primordial counts as metal-enriched. The metal-enriched volume fraction increases monotonically with time and the results from \asloth agree well with most other models in literature.

\begin{figure}
\includegraphics[width=\columnwidth]{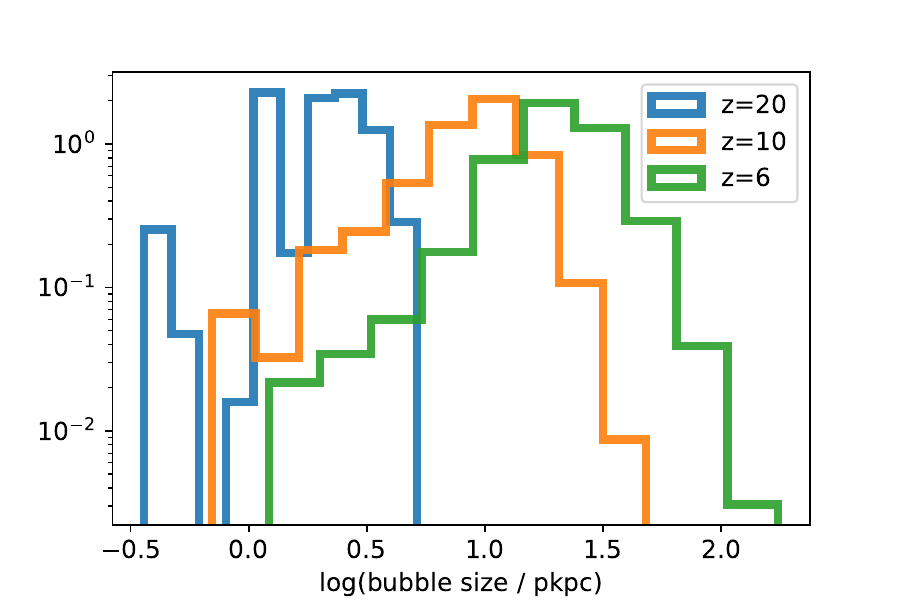}
\caption{Normalized bubble size distribution of metal-enriched bubbles at different redshift. As in Figure~\ref{fig:Rion}, the sizes are shown in physical units.}
\label{fig:Renriched}
\end{figure}

In Fig~\ref{fig:Renriched}, we also show the radius distribution of the metal-enriched bubbles.
The bubbles are created around every halo that experienced chemical feedback. Their radius increases over time and hence these distributions shift to larger bubble sizes with decreasing redshift.
As with the ionized bubbles, the values shown here do not account for the merging of bubbles at low redshifts.

\subsubsection{Outflow Efficiency}
\begin{figure}
\includegraphics[width=\columnwidth]{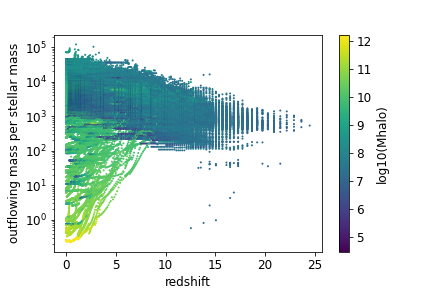}
\caption{Cumulative outflowing mass per cumulative stellar mass for galaxies in the cosmological box at different redshifts. The colorbar indicates the halo mass. In low-mass halos, one stellar baryon can eject $>10^4$ baryons out of the virial radius of a galaxy via radiative and mechanical feedback. In more massive halos, such as the MW, this ratio is closer to unity.}
\label{fig:outflow}
\end{figure}

The sophisticated subgrid model for stellar feedback in \asloth tracks the amount of gas escaping from galaxies. To quantify the outflow efficiencies, we show the outflowing mass per stellar mass in Fig.~\ref{fig:outflow}.
The values shown are cumulative ones, plotted as a function of redshift, and colour-coded by the halo mass. They are equivalent to an SFR-weighted average of the outflow mass-loading factor, $\eta = \dot{M}_{\rm out} / \dot{M}_{*}$.

At high redshift, the halo population is dominated by minihalos and low-mass atomic cooling halos. Outflows from these halos have very high mass loadings, of the order of $10^{3}$ or more, owing to the low gravitational binding energy of these systems, which allows feedback from one or a few massive stars to expel the majority of the gas from the halo. Below a redshift $z \sim 8$, however, we start to form an increasing number of more massive halos. These systems have much lower mass loadings, ranging from $\sim 100$ for systems corresponding to present-day dwarf galaxies, to values $< 1$ for MW-like halos. These values are in reasonable agreement with the results of detailed cosmological simulations of galaxy formation, such as Illustris-TNG \citep{Nelson2019} or FIRE-2 \citep{Pandya2021}.

\section{Caveats}
\label{sec:caveats}
There are several limitations in the current version of \asloth that are important to bear in mind when considering our results.

First, rather than computing the value of the extragalactic LW background based on the SFRD computed by \asloth, we instead use a parameterisation taken from \citet{GreifBromm06}. As discussed in Paper I, the rationale for this choice is that the volume of the Universe contributing to the LW background is typically very large \citep{Ahn2009} and it will often be the case that the evolution of the SFRD in the volume modelled by \asloth is not representative of the evolution of the SFRD in this much larger volume (e.g.\ we expect this to be the case for the Caterpillar merger trees, since these are, by design, not sampling unbiased cosmological volumes). Nevertheless, this choice means that the impact of the LW background is decoupled from out parameter choices for the Pop III IMF and the Pop III and Pop II star formation efficiencies. We plan to relax this limitation in future work, and to investigate the impact of adopting different LW background models \citep[e.g.][]{incatasciato23}.


Second, we use merger trees from a cosmological simulation with side length 8\,Mpc/h for cosmologically representative observables. While this volume is big enough to simulate variations of the SFRD and metal enrichment, it might not be sufficiently big to follow all modes of cosmic reionization and LW feedback \citep{Magg21b}. The limitation here is purely a practical one: as of the writing of this paper, this is the largest simulation that is available to us that has sufficient resolution to model the formation and mergers of minihalos in addition to more massive systems. Should a bigger simulation with similar mass resolution become available in the future, there would be no significant difficulties involved in using it as input for \asloth.

Third, we assume the the Pop~III IMF follows a power-law. While this is motivated by the power-low slope of the present-day IMF over a wide range of masses, this assumption breaks down at both extremes: at low masses, there might be some gradual turnover, and there is also an upper limit on the masses of Pop III stars than can be formed in typical minihalos. Hence, although our derived IMF shape provides solid guidance, other analytical shapes can not be excluded. 

Forth, the recent paper by \citet{koutsouridou23} showed that by assuming the Pop III yields by \citet{HegerWoosley2010}, and thus considering also an unknown energy distribution function for Pop~III stars with masses 10-100 Msun, it is more difficult to limit the Pop III IMF, since there is a degeneracy between the unknown energy distribution function of the first SNe and the unknown Pop IIII IMF.

Finally, we note that after finishing the calibration of the model, we found an inconsistency in the definition of cooling time scales. Hot halo gas cools on the cooling time scale. However, the hot gas' ability to cool is limited by the dynamical time scale. In the previous version of the code, the cooling time could become arbitrarily small, which was fixed in the public version of the code (see commit 026877a6). Now, we only use the cooling time if it is longer than the dynamical time. While this fix has a small effect on star formation at high redshift, it has a negligible effect on the observables used for calibration: 8 of 9 observables match even slightly better with the new code. Only the predicted value for the optical depth worsened slightly from $\tau _\mathrm{old} = 0.044$ to $\tau _\mathrm{new}=0.06179$. Overall, the log-likelihood improved from $\mathcal{L} _\mathrm{old} = -30.09$ to $\mathcal{L} _\mathrm{new} = -29.34$. Because the effect on the observables is minor, we decided not to rerun the expensive calibration. Instead, we updated all figures with results from the correct and current version of the code (only Fig. 4 and 5 are based on the previous results, as they are calibration-specific).

\section{Summary}
This research employed the semi-analytical model \asloth (Ancient Stars and Local Observables by Tracing Halos) to investigate star formation in the early Universe, particularly focusing on the first generation of stars. Given the difficulties in directly observing Pop~III stars, we relied on simulations and indirect observational data to infer their characteristics.

We calibrated \asloth using a likelihood function informed by nine distinct observables, covering both Milky Way-specific and cosmologically representative variables. This thorough calibration ensures that \asloth provides an accurate portrayal of star formation processes in the early Universe, aligning simulated outcomes with observed data across a broad spectrum of parameters. Through this calibration, we derived best-fit values and uncertainties for 11 key parameters critical to understanding early star formation, such as the IMF of Pop~III stars and the escape fractions of ionizing photons. In our best fit model, the Pop III IMF is top heavy, with a high minimum mass ($M_{\rm min} > 10 \: {\rm M_{\odot}}$) and a slope that is shallower than that of the Salpeter IMF, albeit somewhat steeper than the log-flat IMF predicted by simulations. Using a similar approach, \citet{koutsouridou24} showed that the non-detection of mono-enriched PISNe descendants in the Galactic halo can exclude Pop III IMF slopes $<0.75$ if the IMF peak is at masses $>10\Msun$ (see their Fig.1). We also find a high value for the Pop III star formation efficiency per free-fall time, which we interpret as being primarily due to the short duration of Pop III star formation in minihalos relative to the halo free-fall time. 
However, our results show that many of these constrained parameters have large uncertainty intervals, highlighting the difficulties involved in constraining the nature of the first stars based on current observations. 

Finally, we also show some examples of \asloth data products, such as the recovery time distribution, and the size distribution of ionized and metal-enriched bubbles. Comparison of these distributions with the results of detailed numerical simulations provides additional constraints on the nature of Pop III stars.


In summary, \asloth has proven to be an effective tool for simulating and elucidating the complex dynamics of early Universe star formation. This research bridges the gap between theoretical models and observational data, contributing significantly to our knowledge of the primordial epochs of star formation. In a future study, we plan to look in more detail into the baryonic physics that governs the formation of the first generations of stars, provide further useful data products from \asloth, and provide guidance to design future observations to better understand the nature of the first stars.

\section*{Acknowledgements}
We acknowledge funding from JSPS KAKENHI Grant Numbers 19K23437 and 20K14464. We thank Miho Ishigaki and Alex Ji for their valuable discussions and the reviewer for helpful feedback. We acknowledge financial support from the German Excellence Strategy via the Heidelberg Cluster of Excellence (EXC 2181 - 390900948) ``STRUCTURES''. We also acknowledge computing resources provided by the Ministry of Science, Research and the Arts (MWK) of the State of Baden-W\"{u}rttemberg through bwHPC and the German Science Foundation (DFG) through grants INST 35/1134-1 FUGG and 35/1597-1 FUGG, and also data storage at SDS@hd funded through grants INST 35/1314-1 FUGG and INST 35/1503-1 FUGG.

\section*{Data Availability}
The underlying data will be shared upon reasonable request by the authors. The results of the MCMC exploration are available as online-only material. The source code of \asloth\ is available online\footnote{\url{https://gitlab.com/thartwig/asloth}}.



\bibliographystyle{mnras}
\bibliography{MyBib} 



\appendix

\section{Carbon footprint of simulations}
The main MCMC exploration ran on $\sim 100$ Intel Xeon Gold 6230 CPUs for about 6\,weeks. This equals to about 650kWh. Assuming 0.22\,kg of CO$_2$ per kWh for Japan, these calculations have created $\sim 150$\,kg of CO$_2$.


\bsp	
\label{lastpage}
\end{document}